\documentclass{ws-mpla}

\usepackage{color}
\usepackage[hidelinks]{hyperref}

\begin{document}

\renewcommand{\draftnote}{}
\renewcommand{\trimmarks}{}

\markboth{E. A. Matute} {Neutrino flavor mixing with approximate
$\mu$-$\tau$ symmetry}

\catchline{}{}{}{}{}

\title{Neutrino flavor mixing with approximate $\mu$-$\tau$ symmetry \\
within the low-scale minimal linear seesaw model}

\author{\footnotesize Ernesto A. Matute}

\address{Departamento de F\'{\i}sica, Facultad de Ciencia, \\
Universidad de Santiago de Chile (USACH), Santiago, Chile\\
\href{mailto:ernesto.matute@usach.cl}{ernesto.matute@usach.cl}}

\maketitle

\pub{}{}

\begin{abstract}

Neutrino flavor mixing is explained within the recently proposed low-scale
minimal linear seesaw model of neutrino mass generation, a variant of the
standard linear seesaw led by a Dirac pair of extra sterile neutrinos which
is odd under a discrete $Z_2$ symmetry and has a mass at or below the GeV
scale. The tri-bimaximal mixing and its deviations are derived
straightforwardly on the basis of the symmetry between $\mu$ and $\tau$
flavors, without introducing non-abelian discrete flavor symmetries in
the first place.

\end{abstract}

\ccode{PACS Nos.: 14.60.St, 14.60.Pq, 14.60.Lm, 13.15.+g}

\noindent \\[3pt]
The origin of the tiny mass of neutrinos and their large mixing angles are
among the most striking puzzles that remain in particle physics and
cosmology. These phenomena have been verified by noticeable solar,
atmospherical, reactor, and long-baseline neutrino experiments.\cite{PDG}
In connection with the numerous schemes that have been proposed over the
years to understand them, the minimal seesaw model,\cite{minimalseesaw}
a simplified variant of the classical type-I seesaw case with just two sterile
right-handed (RH) neutrinos,\cite{Seesaw1}$^{\mbox{--}}$\cite{Seesaw5} appears
as an economical and very predictive one that has been analyzed from many
points of view. One of its most important features is that the lightest of
the three active neutrinos becomes massless, which fixes the mass scale of
the other two after neutrino oscillation data on neutrino mass squared
differences are taken into account.

Our aim in this paper is to explain for the first time the neutrino mixing
in terms of the tri-bimaximal pattern\cite{TBM} and its deviations,
within an alternative model proposed recently to motivate the sub-GeV scale for
sterile neutrinos,\cite{EAM} in which none of the RH neutrinos of the minimal seesaw
model correspond to those that generate the neutrino masses. This assumption is
suggested by the absence of conclusive evidence of effects from RH neutrinos at low
energies. The new model is an extension of the standard model (SM) with three RH
neutrinos ($\nu_R$) and a Dirac pair of extra sterile neutrinos ($N_R,N_L$), which are
distinguished among themselves by their assignments to an additional discrete $Z_2$
symmetry. The RH neutrinos, and all of the fermions of the SM, are in the even state,
while the extra sterile neutrinos are in the odd state. This establishes a left--right
symmetry in the neutrino content of the $Z_2$ even sector that is not broken by the
extra neutrinos in the odd state. The canonical high-scale seesaw is used to suppress
all of the mass terms involving RH neutrinos. The extra sterile neutrinos lead a
seesaw mechanism of neutrino mass generation at or below the GeV scale, which is a
low-scale variant of the well-known linear
seesaw\cite{Linear1}$^{\mbox{--}}$\cite{Linear3} and its simplified minimal version
with two RH neutrinos.\cite{MLSS1}$^{\mbox{--}}$\cite{MLSS3}

The model should also be contrasted with those more economical ones where just three
RH neutrinos are included, but having a severe segregation between two of them, heavy
enough to be responsible for the light neutrino masses, and the lightest third one,
accommodated as a dark matter (DM) candidate.\cite{Shapo1}$^{\mbox{--}}$\cite{India}
In our case, the discrimination, based on the $Z_2$ symmetry, is instead between the three
heavy RH neutrinos and the extra sterile neutrinos, in charge of neutrino mass generation
and also seen as DM candidates. A clear difference is that in the former, the lightest RH
neutrino is of Majorana type, whereas in the latter, the extra neutrinos build a quasi-Dirac
neutrino.

Sterile neutrinos at the sub-GeV scale, on the other hand, have been considered in
different scenarios by several authors, for instance, see
Refs.~\refcite{Shapo1}--\refcite{Model8}.
Constraints from laboratory, astrophysical and cosmological observations vary according to
the mass and mixing of such heavy neutral leptons.\cite{GeVne1}$^{\mbox{--}}$\cite{GeVne4}
But, in these works, either RH Majorana neutrinos are included or their extensions with extra
sterile neutrinos, which have no influence on the genesis of active neutrino masses, are
discussed. Although these approaches have interesting experimental signatures, they are
different from our proposition from the standpoint of the model as well as the phenomenology.
Our sub-GeV seesaw model implies no low-energy phenomenology with ordinary RH neutrinos, in
contrast to the alternative low-scale seesaw models that require them to generate mass and
mixing of light neutrinos. These RH neutrinos are the ones that would be responsible for the
baryon asymmetry in the universe via leptogenesis\cite{leptogenesis} and charged under an
eventual $\mbox{U}(1)_{B-L}$ gauge symmetry and a possible $\mbox{SU}(2)_L \times \mbox{SU}(2)_R
\times \mbox{U}(1)_{B-L}$ gauge extension of the SM,\cite{LRsym1}$^{\mbox{--}}$\cite{LRsym3} just
the opposite to our sterile neutrinos which would remain singlets in such scenario. Our proposal
would be ruled out if signals of RH weak currents are found at low energies. Also, the extra sterile
neutrinos, being of quasi-Dirac-type, practically have no contribution to lepton number violating
processes such as neutrinoless double beta decays. Still, they can mediate rare lepton flavor
violating transitions like $\mu \rightarrow e \gamma$. Nonetheless the active-sterile neutrino mixing
angles are not large enough to be at the range of the current experimental sensitivity.\cite{MEG}
Since there is no tree-level coupling between sterile neutrinos and nucleons, the model also
avoids the strong constraints from direct detection.\cite{xenon} However, they can be explored at
the collider facilities through two-body and three-body
decays.\cite{Collider1}$^{\mbox{--}}$\cite{Collider5}

Our starting point is the effective $8 \times 8$ neutrino mass matrix in the basis
($\nu_L, \nu^c_R, N^c_R, N_L$) which after the electroweak symmetry breaking is given by
\begin{equation}
\mathcal{M}_\nu = \left(
\begin{array}{ccccccc}
0 && m_D && m^\prime_D && \mu^\prime_L \\[3pt] m^T_D && m_R && \mu^\prime_R
&& \mu^\prime_D \\[3pt] m^{\prime T}_D && \mu^{\prime T}_R && 0 && M_D \\[3pt]
\mu^{\prime T}_L && \mu^{\prime T}_D && M_D && 0
\end{array} \right) ,
\label{massmatrixDM}
\end{equation}
where a basis in which the mass matrix of charged leptons appears real and
diagonal has been taken and Majorana mass terms of extra sterile neutrinos
have been excluded.\cite{EAM}

More specifically, we circumscribe our analysis to the low-scale region in which the
heavy RH neutrinos are essentially decoupled from active and extra sterile neutrinos
via the classic high-scale seesaw, suppressing the mass terms $m_D$, $\mu^\prime_R$,
and $\mu^\prime_D$. It is assumed that they do not influence phenomenology at low
energies, in agreement with the null experimental evidence so far proving the
existence of RH neutrinos. In the basis ($\nu_L, N^c_R, N_L$), the effective
$5 \times 5$ mass matrix becomes
\begin{equation}
\mathcal{M}^\prime_{\nu} = \left(
\begin{array}{ccccc}
0 && m^\prime_D && \mu^\prime_L \\[3pt] m^{\prime T}_D && 0 && M_D \\[3pt]
\mu^{\prime T}_L && M_D && 0
\end{array} \right) .
\label{effmassmatrix}
\end{equation}
The mass hierarchy is $m^\prime_D \sim \mu^\prime_L \ll M_D$, justified
by the $Z_2$ symmetry between the active neutrinos and the extra sterile
neutrinos, softly broken by the mass terms $m^{\prime}_D \overline{\nu_L} N_R$
and $\mu^{\prime}_L \overline{\nu_L} N^c_L$. We explain below why the
second and third diagonal elements of this mass matrix are taken to be zero.

The origin and soft breaking of the $Z_2$ symmetry can be understood by
turning the model to the so-called presymmetry presented in Ref.~\refcite{Presym}.
Presymmetry identifies an underlying electroweak theory of leptons and quarks based
on the requirements of left--right symmetry in fermionic content and symmetric fractional
electroweak charges. It is built upon the global $\mbox{U}(1)_{B-L}$ symmetry, which rules
out Majorana mass terms. Presymmetry is broken at the stage of leptons and quarks,
allowing symmetry breaking terms such as the Majorana mass terms for RH neutrinos
and those corresponding to the $Z_2$ symmetry. In a sense, presymmetry manifests
the high affinity of the active left-handed (LH) neutrinos with their RH partners
prior to the breaking that separates them. It is the source of the $Z_2$ symmetry
through which the new leveling of singlet fermions is introduced. Presymmetry also
explains why the extra sterile neutrinos cannot have Majorana mass terms and why
the three RH neutrinos are required in spite of being so heavy that they do not
influence phenomenology at low energies. Presymmetry breaking changes couplings of
LH and RH neutrinos, but Majorana mass terms for the extra sterile neutrinos
(second and third diagonal elements of $\mathcal{M}^\prime_{\nu}$ in
Eq.~(\ref{effmassmatrix})) remain blocked.

The effective light neutrino mass, following the diagonalization of the mass \linebreak
matrix in Eq.~(\ref{effmassmatrix}), is approximately given by
\begin{equation}
m_\nu = - \frac{m^\prime_D \mu^{\prime T}_L}{M_D} -
\frac{\mu^\prime_L m^{\prime T}_D}{M_D} ,
\label{light}
\end{equation}
while for the heavier neutrino masses we have
\begin{equation}
\begin{array}{l}
\displaystyle M^+_N = M_D + \frac{(m^\prime_D + \mu^\prime_L)^T
(m^\prime_D + \mu^\prime_L)}{2 M_D} , \\ \\
\displaystyle M^-_N = - M_D - \frac{(m^\prime_D - \mu^\prime_L)^T
(m^\prime_D - \mu^\prime_L)}{2 M_D} .
\end{array}
\label{heavy}
\end{equation}
Note that the null trace of the mass matrix $\mathcal{M}^\prime_\nu$
implies
\begin{equation}
tr(m_\nu) + M^+_N + M^-_N =0 ,
\end{equation}
so that the mass splitting of the pseudo-Dirac neutrino is equal to the
sum of active neutrino masses and the Dirac pair of sterile neutrinos can
therefore be treated effectively as a Dirac neutrino.

The mass matrix in Eq.~(\ref{effmassmatrix}) and its eigenvalues in
Eqs.~(\ref{light}) and (\ref{heavy}) are similar in form to the ones
discussed in the linear seesaw. Moreover, one might argue that the whole
process of going with the extra sterile neutrinos while keeping decoupled
the RH neutrinos is a mere relabeling of the linear seesaw, though with
additional decoupled gauge-singlet neutrinos. Specifically, with the
notation change $N_R \rightarrow \nu_R$, $N_L \rightarrow S_L$, and
$m^\prime_D \rightarrow m_D$, one would have a neutrino mass matrix that
in the basis ($\nu_L, \nu^c_R, S_L$) reads as
\begin{equation}
\mathcal{M}^\prime_{\nu} = \left(
\begin{array}{ccccc}
0 && m_D && \mu^\prime_L \\[3pt] m^T_D && 0 && M_D \\[3pt]
\mu^{\prime T}_L && M_D && 0
\end{array} \right) ,
\label{linearmatrix}
\end{equation}
with the light neutrino mass given by
\begin{equation}
m_\nu = - m_D (\mu^\prime_L M^{-1}_D)^T - \mu^\prime_L M^{-1}_D m^T_D
\label{Nemass3}
\end{equation}
that is proper of the linear seesaw scheme. However, when we compare the
mass terms $m_D \overline{\nu_L} \nu_R$ and $m^\prime_D \overline{\nu_L} N_R$,
for instance, we cannot conclude that this involves just a change of notation:
there is invariance under $Z_2$ symmetry transformations in the first term,
but not in the second one. Besides, $m^\prime_D \ll m_D$ under the soft breaking
of $Z_2$. This implies that the mass $M_D$ of the heavy Dirac neutrino in the
new model is much lower than that in the usual one.

On the other hand, from the phenomenological point of view, there exists an extra
suppression with respect to the usual linear seesaw which essentially is in accord with
\begin{equation}
m_\nu = \frac{2 m_D \mu_L^\prime}{M_D} \cdot \frac{m_D^\prime}{m_D} ,
\end{equation}
leading to $M_D \sim 1$ GeV for $m_\nu \sim 10^{-2}$ eV and
$\mu_L^\prime \sim m_D^\prime \sim 4$ keV, associated with the soft breaking
of the $Z_2$ symmetry. That is, $m_D^\prime \sim 10^{-5} \, m_D$ for
$m_D \approx m_\tau \sim 1$~GeV, attributed to the similarity condition between
the Dirac neutrino mass and the Dirac mass of the charged lepton for the third generation of
neutrinos.\cite{EAM} Therefore, a seesaw scale ($M_D$) about five orders of magnitude smaller
than the scale of the usual linear seesaw is implied. This can be diminished to
$m_D^\prime \sim 10^{-2} \, m_D$ if the similarity condition involving the first generation
of charged leptons is used instead, but with the $Z_2$ symmetry breaking terms maintained
at the keV range, same as done for the inverse and linear seesaw mechanisms, making the model
more appealing and emphasizing its differences. As argued above, the soft breaking of the
$Z_2$ symmetry would have its origin in the presymmetry breaking, allowing the coupling of LH
and RH neutrinos to the extra sterile neutrinos in such a way that the new variant has a
seesaw scale at the GeV range independently of the similarity condition used between
the size of the Dirac neutrino mass ($m_D$) and the Dirac mass of the charged lepton, marking
then a difference with the other mechanisms. Note that the scale of $M_D$ can be brought down
even to the keV scale for $\mu_L^\prime$ and $m_D^\prime$ at the eV range.

The claim is that this model, proposed to understand the generation of light neutrino masses,
is a new low-scale minimal variant of the standard linear seesaw, led by the extra sterile
neutrinos in the $Z_2$ odd state instead of the RH neutrinos in the even state.

We now proceed with the analysis of neutrino flavor mixings in the framework of the above model.
The sterile neutrinos become approximately the heavy mass eigenstates, while the active neutrino
mass matrix, diagonalized by a unitary matrix $U$ as
\begin{equation}
U^T m_\nu U = m^{\rm diag}_\nu = \mbox{diag} (m_1, m_2, m_3) ,
\label{diag}
\end{equation}
is the one that leads us to determine the light mass eigenstates $\nu_i$
($i=1,2,3$) and find approximately the mixing of active neutrinos among
themselves by means of
\begin{equation}
\nu_\alpha = U_{\alpha i} \nu_i ,
\label{mixing}
\end{equation}
where $\nu_\alpha$~($\alpha = e, \mu, \tau$) denotes the flavor eigenstates.
The tiny mixing between active and sterile neutrinos, given by
$U_{\alpha N} \sim \mu^\prime_{L\alpha} / M_D \lesssim 10^{-3}$, allows us to
use the unitarity property of matrix $U$ in Eq.~(\ref{mixing}), neglecting
thus its departure from unitarity. Actually, it is unitary at the
$\mathcal{O}(10^{-2})$ level according to current neutrino oscillation data
and electroweak precision measurements.\cite{Unitary} Furthermore, since the
CP violating phases of the unitary matrix $U$ are not measured with precision
in present neutrino experiments,\cite{CP} we restrict ourselves by taking
them to be zero. That is, in view of the fact that the CP violating
phases are currently unknown\cite{CPphase} and that the CP-conserving regime is
still admitted, here we stick to the case of $U$ being real and work with a focus
on the tri-bimaximal mixing and its deviations. A discussion on how to accommodate
eventual CP violating phases implies to consider $U$ as a complex matrix, which is
beyond the scope of this paper.

Using the notation of Ref.~\refcite{Shapo1}, making positive the mass terms in
Eq.~(\ref{light}) by phase redefinition in chiral fields, the diagonalized
matrix of active neutrinos can be rewritten in the following form:
\begin{equation}
m^{\rm diag}_\nu = S + S^T ,
\label{dsmatrix}
\end{equation}
where $S$ is defined as
\begin{equation}
S_{ij} = X_i Y_j
\end{equation}
with
\begin{equation}
X_i = \frac{1}{\sqrt{M_D}} (m^{\prime T}_D U)_i , \quad
Y_i = \frac{1}{\sqrt{M_D}} (\mu^{\prime T}_L U)_i .
\label{XY}
\end{equation}
By taking the trace of both sides of Eq.~(\ref{dsmatrix}), it is
found that
\begin{equation}
m_1+m_2+m_3 = \sum_{i=1}^3 2 \, X_i Y_i .
\end{equation}

Now, from the equations $\mbox{det}(S)=0$ and $\mbox{det}(S+S^T)=0$,
it is deduced that the lightest active neutrino is massless:
$m_1=0$, considering the normal ordering $m_1 < m_2 < m_3$ of neutrino
masses (or $m_3=0$ in the case of the inverted ordering $m_3 < m_1 < m_2$,
changing subindexes 1, 2, 3 by 3, 1, 2, respectively, in everything that
follows below). Accordingly, it is found that
\begin{equation}
m_2 m_3 \, X_1 Y_1 = 0 ,
\end{equation}
which demands $X_1 Y_1 = 0$. Actually, this means that
\begin{equation}
X_1 =0 , \quad Y_1 = 0 ,
\end{equation}
since one implies the other as the off-diagonal elements of
the matrix $S+S^T$ must vanish. Thus, Eq.~(\ref{dsmatrix})
is reduced to that for $2 \times 2$ matrices:
\begin{equation}
\left( \begin{array}{ccc}
m_2 && 0 \\ 0 && m_3
\end{array} \right) =
\left( \begin{array}{ccc}
X_2 Y_2 && X_2 Y_3 \\ X_3 Y_2 && X_3 Y_3
\end{array} \right) +
\left( \begin{array}{ccc}
X_2 Y_2 && X_3 Y_2 \\ X_2 Y_3 && X_3 Y_3
\end{array} \right) .
\label{22matrix}
\end{equation}
Specifically,
\begin{equation}
X_2 Y_2 = \frac{m_2}{2} , \quad X_3 Y_3 = \frac{m_3}{2} , \quad X_3 Y_2 = - X_2 Y_3 ,
\end{equation}
which imply
\begin{equation}
X_2 Y_3 = i \frac{\sqrt{m_2 m_3}}{2} .
\end{equation}

The vanishing determinant of the full mass matrix means no additional
constraints on these mixing couplings. We choose $Y_2=X_2$, which leads to
\begin{equation}
Y_2 = X_2 = \sqrt{\frac{m_2}{2}} , \quad Y_3 = X^{*}_3 = i \sqrt{\frac{m_3}{2}} \, .
\label{XYvalues}
\end{equation}

By using Eqs.~(\ref{XY}) and (\ref{XYvalues}), the following sum rule is obtained:
\begin{equation}
\frac{\sum_\alpha (|\mu^\prime_{L\alpha}|^2+|m^\prime_{D\alpha}|^2)}{\sum_i (|X_i|^2+|Y_i|^2)}
= \frac{\sum_\alpha (|\mu^\prime_{L\alpha}|^2+|m^\prime_{D\alpha}|^2)}{\sum_i m_i} = M_D .
\label{sumrule}
\end{equation}

On the other hand, from Eq.~(\ref{XY}), the expressions of $m^\prime_D$ and
$\mu^\prime_L$ are found to be
\begin{equation}
\begin{array}{l}
\displaystyle \frac{m^\prime_{D\alpha}}{\sqrt{M_D}} = X_2 U_{\alpha 2} + X_3 U_{\alpha 3} , \\ \\
\displaystyle \frac{\mu^\prime_{L\alpha}}{\sqrt{M_D}} = Y_2 U_{\alpha 2} + Y_3 U_{\alpha 3} .
\end{array}
\end{equation}
With $U$ taken to be real, it is seen that
\begin{equation}
m^\prime_{D\alpha} = \mu^{\prime *}_{L\alpha} .
\end{equation}
Hence, $|m^\prime_D| = |\mu^\prime_L|$, which is a characteristic of the new
variant of the usual linear seesaw model. As commented above, in the standard
scheme, with $m_D$ instead of $m^\prime_D$, the result is $|m_D| \gg |\mu^\prime_L|$
and therefore a much higher seesaw scale.

No further conditions are established on the mass parameters, so that any given
ansatz will fix the elements of the matrix $U$. We enforce constraints as
dictated by a symmetry between the $\mu$ and $\tau$ flavors of the second and
third generations of neutrinos:
\begin{equation}
\begin{array}{l}
m^\prime_{D\mu} = \mu^{\prime}_{L\tau} , \\ \\
m^\prime_{De} = (\mu^\prime_{L\mu} + \mu^\prime_{L\tau}) / 2 ,
\end{array}
\label{ansatz}
\end{equation}
although the imaginary part of the second condition is redundant. These,
together with the unitary restraints, lead straightforwardly to the
tri-bimaximal (TBM) matrix\cite{TBM}
\begin{equation}
U_{\mathsf{TBM}} = \left( \begin{array}{ccc}
\displaystyle \frac{2}{\sqrt{6}} & \;\; \displaystyle \frac{1}{\sqrt{3}} \;\; & 0 \\ && \\
\displaystyle -\frac{1}{\sqrt{6}} & \displaystyle \frac{1}{\sqrt{3}} &
\displaystyle -\frac{1}{\sqrt{2}} \\ && \\
\displaystyle -\frac{1}{\sqrt{6}} & \displaystyle \frac{1}{\sqrt{3}} &
\displaystyle \frac{1}{\sqrt{2}}
\end{array} \right)
\end{equation}
and the following expressions for $m^\prime_D$ and $\mu^\prime_L$:
\begin{equation}
\begin{array}{l}
\displaystyle \frac{m^\prime_{De}}{\sqrt{M_D}} = \sqrt{\frac{m_2}{6}} =
\frac{\mu^\prime_{Le}}{\sqrt{M_D}} , \\ \\
\displaystyle \frac{m^\prime_{D\mu}}{\sqrt{M_D}} = \sqrt{\frac{m_2}{6}} +
i \frac{\sqrt{m_3}}{2} = \frac{\mu^{\prime *}_{L\mu}}{\sqrt{M_D}} , \\ \\
\displaystyle \frac{m^\prime_{D\tau}}{\sqrt{M_D}} = \sqrt{\frac{m_2}{6}} -
i \frac{\sqrt{m_3}}{2} = \frac{\mu^{\prime *}_{L\tau}}{\sqrt{M_D}} .
\end{array}
\label{TBMmixings}
\end{equation}

In this manner, with the $\mu$-$\tau$ symmetric entries in Eq.~(\ref{ansatz}), we are
incorporating into the model the $\mu$-$\tau$ symmetry ($|U_{\mu i}|=|U_{\tau i}|$),\cite{mu-tau}
which in our scheme means
$|m^\prime_{D\mu}|=|m^\prime_{D\tau}|=|\mu^\prime_{L\mu}|=|\mu^\prime_{L\tau}|$,
and the zero value for the $U_{e3}$ matrix element. For a review about the $\mu$-$\tau$
flavor symmetry, see Ref.~\refcite{FlavorSym}.

We note that
\begin{equation}
|m^\prime_{De}|^2 = \frac{m_2}{m_2+3 m_3 /2} \, |m^\prime_{D\mu}|^2 .
\label{mDe}
\end{equation}
Thus, by combining Eqs.~(\ref{sumrule}) and (\ref{mDe}), and using the values
of neutrino masses in the normal ordering\cite{PDG} ($m_i=(0, 0.0086, 0.05)$~eV) with
$|m^\prime_{D\mu}| = |m^\prime_{D\tau}| \sim \mbox{4 keV}$, we have
$|m^\prime_{De}| \sim \mbox{1 keV}$ and $M_D \sim \mbox{1 GeV}$ (or about
2 keV and 1 GeV, respectively, in the case of the inverted ordering
($m_i=(0.0492, 0.05, 0)$~eV)), i.e. as already mentioned above, light neutrino masses
could be generated from new physics around the GeV level, accessible at collider facilities.
Similarly, for $|m^\prime_{D\mu}|$ and $|m^\prime_{D\tau}|$ now at the eV range, the mass of
the sterile neutrino is set at the keV scale, within the reach of X-ray searches.\cite{Xray}

The conventional parametrization of the unitary matrix,\cite{PDG} in contrast,
adopts the form of
\begin{equation}
U = R_{23} R_{13} R_{12},
\label{usedU}
\end{equation}
where
\begin{equation}
R_{23} = \left(
\begin{array}{ccc}
1 & 0 & 0 \\ 0 & c_{23} & s_{23} \\ 0 & -s_{23} & c_{23}
\end{array} \right) , \quad
R_{13} = \left(
\begin{array}{ccc}
c_{13} & 0 & s_{13} \\ 0 & 1 & 0 \\ -s_{13} & 0 & c_{13}
\end{array} \right) , \quad
R_{12} = \left(
\begin{array}{ccc}
c_{12} & s_{12} & 0 \\ -s_{12} & c_{12} & 0 \\ 0 & 0 & 1
\end{array} \right) ,
\label{Rmatrices}
\end{equation}
with $s_{ij}=\sin \theta_{ij}$ and $c_{ij}=\cos \theta_{ij}$ ($ij=23,13,12$).
When compared with our way of doing for the TBM mixing, we find that
\begin{equation}
\begin{array}{ll}
\displaystyle c_{23} = \frac{m^\prime_{D\tau}-m^\prime_{D\mu}}{2X_3\sqrt{M_D}} = \frac{1}{\sqrt{2}} , & \quad
\displaystyle s_{23} = \frac{m^\prime_{D\mu}-m^\prime_{D\tau}}{2X_3\sqrt{M_D}} = -\frac{1}{\sqrt{2}} , \\ \\
\displaystyle c_{13} = \frac{m^\prime_{D\tau}-m^\prime_{D\mu}}{\sqrt{2}X_3\sqrt{M_D}} = 1 , & \quad
\displaystyle s_{13} = \frac{m^\prime_{De}-m^{\prime *}_{De}}{2X_3\sqrt{M_D}} = 0 , \\ \\
\displaystyle c_{12} = \frac{m^\prime_{D\tau}+m^\prime_{D\mu}}{\sqrt{2}X_2\sqrt{M_D}} = \frac{2}{\sqrt{6}} , & \quad
\displaystyle s_{12} = \frac{m^\prime_{De}+m^{\prime *}_{De}}{2X_2\sqrt{M_D}} = \frac{1}{\sqrt{3}} .
\end{array}
\label{standardangles}
\end{equation}
These expressions show us the connections between the three standard neutrino flavor
mixing angles and the mixing parameters referred to the interactions of the extra
sterile neutrinos in the $Z_2$ odd state. Interestingly, they define the correlations
between neutrino mass ratios and flavor mixing angles.

As far as the light neutrino mass matrix is concerned, from Eqs.~(\ref{light}) and
(\ref{TBMmixings}) we obtain
\begin{equation}
m_\nu = \left( \begin{array}{ccc}
\displaystyle \frac{m_2}{3} & \displaystyle \frac{m_2}{3} &
\displaystyle \frac{m_2}{3} \\ && \\
\displaystyle \frac{m_2}{3} & \;\;\; \displaystyle \frac{m_2}{3}+\frac{m_ 3}{2} \;\;\; &
\displaystyle \frac{m_2}{3}-\frac{m_3}{2} \\ && \\
\displaystyle \frac{m_2}{3} & \displaystyle \frac{m_2}{3}-\frac{m_3}{2} &
\displaystyle \frac{m_2}{3}+\frac{m_3}{2}
\end{array} \right) ,
\label{massTBMa}
\end{equation}
which can be rewritten as
\begin{equation}
m_\nu = \frac{m_3}{2} \left(
\begin{array}{ccccc}
1 && 0 && 0 \\0 && 1 && 0 \\ 0 && 0 && 1
\end{array} \right) -
\frac{m_3}{2} \left(
\begin{array}{ccccc}
1 && 0 && 0 \\0 && 0 && 1 \\ 0 && 1 && 0
\end{array} \right)
+ \; \frac{m_2}{3} \left(
\begin{array}{ccccc}
1 && 1 && 1 \\ 1 && 1 && 1 \\ 1 && 1 && 1
\end{array} \right) .
\label{massTBMb}
\end{equation}

It is worth mentioning here that the mass matrix for TBM mixing in
Eq.~(\ref{massTBMb}) is invariant under the action of the unitary,
real, symmetric matrices $S$ and $G$\cite{TBMsym}:
\begin{equation}
m_\nu = S m_\nu S , \quad m_\nu = G m_\nu G ,
\label{mtrans}
\end{equation}
where
\begin{equation}
S = \frac{1}{3} \left(
\begin{array}{rcrcr}
-1 && 2 && 2 \\ 2 && -1 && 2 \\ 2 && 2 && -1
\end{array} \right) , \quad
G = \left(
\begin{array}{rcrcr}
1 && 0 && 0 \\0 && 0 && 1 \\ 0 && 1 && 0
\end{array} \right) ,
\label{SGmatrices}
\end{equation}
with $S^2=G^2=1$, i.e. matrices of order two.
Moreover, $S$ and $G$ are matrices of the $S_4$ group, while only $S$ is an
element of the $A_4$ group. This is why the $S_4$ and $A_4$ groups have
been employed for TBM mixing, assuming that the theory is invariant under
the symmetry described by these groups.\cite{TBMgroups1,TBMgroups2}
In contrast to this way of model building, our approach to the TBM mixing
relies just on the conditions established in Eq.~(\ref{ansatz}), which
reflect $\mu$-$\tau$ flavor symmetry.

The experimental observation of a nonzero $U_{e3}$, however, demands to change
the above flavor mixing pattern. This means that the exact flavor symmetry in
Eq.~(\ref{ansatz}), though very simple, does not apply, as expected; after all,
exact symmetries are not suitable to fermion masses and flavor mixing. Yet,
neutrino oscillation data provide values so close to $U_{\mathsf{TBM}}$ that
this should be taken as the correct first approximation. An economical possibility
to have a phenomenological viable modified matrix is to maintain the first or
second column of $U_{\mathsf{TBM}}$ unaltered and adjust its other two columns
within the unitarity constraints. Following this line of analysis, we obtain an
appropriate version of $U_{\mathsf{TBM}}$ by amending Eq.~(\ref{ansatz}) with
an explicit breaking term in the following form:
\begin{equation}
\begin{array}{l}
\mbox{Re} \; m^\prime_{D\mu} = \mbox{Re} \; \mu^{\prime}_{L\tau} , \quad
\mbox{Im} \; m^\prime_{D\mu} = (1+\epsilon) \; \mbox{Im} \; \mu^\prime_{L\tau} ,
\\ \\ \mbox{Re} \; m^\prime_{De} = \mbox{Re} \; (\mu^{\prime}_{L\mu}
+ \mu^{\prime}_{L\tau}) / 2 ,
\end{array}
\label{ansatz11}
\end{equation}
with $\epsilon$ small and real, i.e. only a small deviation from the imaginary
part of the first symmetric constraint given in Eq.~(\ref{ansatz}), but leaving
out the unnecessary imaginary part of the second relationship. With this correction
the flavor mixing matrix, up to first order in $\epsilon$, becomes
\begin{equation}
U = \left( \begin{array}{ccc}
\displaystyle \frac{2}{\sqrt{6}} & \displaystyle \frac{1}{\sqrt{3}} &
\displaystyle \frac{\epsilon}{\sqrt{2}} \\ && \\
\displaystyle -\frac{1}{\sqrt{6}} \left( 1-\frac{3\epsilon}{2} \right) &
\;\; \displaystyle \frac{1}{\sqrt{3}} \;\; &
\displaystyle -\frac{1}{\sqrt{2}} \left (1+\frac{\epsilon}{2} \right) \\ && \\
\displaystyle -\frac{1}{\sqrt{6}} \left( 1+\frac{3\epsilon}{2} \right) &
\displaystyle \frac{1}{\sqrt{3}} &
\displaystyle \frac{1}{\sqrt{2}} \left( 1-\frac{\epsilon}{2} \right)
\end{array} \right) ,
\label{Udeviations}
\end{equation}
with the second column elements unchanged, as desired.

The modifications change $m^\prime_D$ and $\mu^\prime_L$ in Eq.~(\ref{TBMmixings})
according to
\begin{equation}
\begin{array}{l}
\displaystyle \frac{m^\prime_{De}}{\sqrt{M_D}} = \sqrt{\frac{m_2}{6}} - i \frac{\sqrt{m_3}}{2} \epsilon =
\frac{\mu^{\prime *}_{Le}}{\sqrt{M_D}} , \\ \\
\displaystyle \frac{m^\prime_{D\mu}}{\sqrt{M_D}} = \sqrt{\frac{m_2}{6}} +
i \frac{\sqrt{m_3}}{2} \left( 1+\frac{\epsilon}{2} \right) = \frac{\mu^{\prime *}_{L\mu}}{\sqrt{M_D}} , \\ \\
\displaystyle \frac{m^\prime_{D\tau}}{\sqrt{M_D}} = \sqrt{\frac{m_2}{6}} -
i \frac{\sqrt{m_3}}{2} \left( 1-\frac{\epsilon}{2} \right) = \frac{\mu^{\prime *}_{L\tau}}{\sqrt{M_D}} .
\end{array}
\label{massdeviations}
\end{equation}
Note that deviations occur only in the imaginary parts of $m^\prime_D$ and
$\mu^\prime_L$. They are the minimal changes to comply with experimental
results.

The revised version of the light neutrino mass matrix is a modified form of
Eq.~(\ref{massTBMa}) that can be expressed as
\begin{equation}
m_\nu + \delta m_\nu = m_\nu + \epsilon \; \frac{m_3}{2}  \left(
\begin{array}{rcrcr}
0 && -1 && 1 \\ -1 && 1 && 0 \\ 1 && 0 && -1
\end{array} \right) .
\label{deltamass}
\end{equation}
In so doing, we arrive at a mass matrix which is not invariant under the
transformations given in Eqs.~(\ref{mtrans}) and (\ref{SGmatrices}).

Concerning the standard mixing angles in Eq.~(\ref{standardangles}),
the only deviations that we get, up to first order in $\epsilon$, are
\begin{equation}
c_{23}=\frac{1}{\sqrt{2}} \left( 1-\frac{\epsilon}{2} \right) , \quad
s_{23}=-\frac{1}{\sqrt{2}} \left( 1+\frac{\epsilon}{2} \right) , \quad
s_{13}=\frac{\epsilon}{\sqrt{2}} .
\label{angledeviations}
\end{equation}
For example, by taking $\epsilon = 0.202$, the numerical result for the
mixing matrix $U$, obtained from Eq.~(\ref{Udeviations}), is
\begin{equation}
| U | = \left(
\begin{array}{ccc}
0.816 & \;\; 0.577 & \;\; 0.143 \\ 0.285 & \;\; 0.577 & \;\; 0.779 \\ 0.532 & \;\; 0.577 & \;\; 0.636
\end{array} \right) ,
\end{equation}
which is consistent with that extracted from current experimental
data.\cite{minimalseesaw}

Different mixing angles can be accommodated by conveniently varying the
deviations from Eq.~(\ref{ansatz}), like we did in Eq.~(\ref{ansatz11}).
We can compare them, for instance, with the case where the deviation in
Eq.~(\ref{ansatz11}) is extended according to
\begin{equation}
\begin{array}{rl}
m^\prime_{D\mu} &= (1+\epsilon) \; \mu^{\prime}_{L\tau} , \\ \\
\mbox{Re} \; m^\prime_{De} &= \mbox{Re} \; (\mu^{\prime}_{L\mu}
+ \mu^{\prime}_{L\tau}) / 2 ,
\end{array}
\label{ansatz111}
\end{equation}
with $\epsilon$ small and real, which seems to be more consistent and realistic.
In fact, instead of the $U$ matrix in Eq.~(\ref{Udeviations}), we obtain
\begin{equation}
U = \left( \begin{array}{ccc}
\displaystyle \frac{2}{\sqrt{6}} & \displaystyle \frac{1}{\sqrt{3}} &
\displaystyle \frac{2\epsilon}{\sqrt{2}} \\ && \\
\displaystyle -\frac{1}{\sqrt{6}} \left( 1-\frac{5\epsilon}{2} \right) &
\;\; \displaystyle \frac{1}{\sqrt{3}} \left( 1+\frac{\epsilon}{2} \right) \;\; &
\displaystyle -\frac{1}{\sqrt{2}} \left( 1+\frac{\epsilon}{2} \right) \\ && \\
\displaystyle -\frac{1}{\sqrt{6}} \left( 1+\frac{5\epsilon}{2} \right) &
\displaystyle \frac{1}{\sqrt{3}} \left( 1-\frac{\epsilon}{2} \right) &
\displaystyle \frac{1}{\sqrt{2}} \left( 1-\frac{\epsilon}{2} \right)
\end{array} \right) ,
\label{Udeviations2}
\end{equation}
which, with $2\epsilon=0.202$, leads to
\begin{equation}
| U | = \left(
\begin{array}{ccc}
0.816 &  \;\; 0.577 & \;\; 0.143 \\ 0.305 & \;\; 0.607 & \;\; 0.743 \\ 0.511 & \;\; 0.548 & \;\; 0.671
\end{array} \right) ,
\end{equation}
with apparently improved magnitudes for the elements of the second and third
lines.

With respect to the deviations pointed out in Eq.~(\ref{angledeviations}),
the only change to consider is that now
\begin{equation}
s_{13}=2\epsilon/\sqrt{2} ,
\end{equation}
whereas Eq.~(\ref{massdeviations}) is modified as follows:
\begin{equation}
\begin{array}{l}
\displaystyle \frac{m^\prime_{De}}{\sqrt{M_D}} = \sqrt{\frac{m_2}{6}} -
i \frac{\sqrt{m_3}}{2} 2\epsilon = \frac{\mu^{\prime *}_{Le}}{\sqrt{M_D}} , \\ \\
\displaystyle \frac{m^\prime_{D\mu}}{\sqrt{M_D}} = \left( \sqrt{\frac{m_2}{6}} +
i \frac{\sqrt{m_3}}{2} \right) \left( 1+\frac{\epsilon}{2} \right) =
\frac{\mu^{\prime *}_{L\mu}}{\sqrt{M_D}} , \\ \\
\displaystyle \frac{m^\prime_{D\tau}}{\sqrt{M_D}} = \left( \sqrt{\frac{m_2}{6}} -
i \frac{\sqrt{m_3}}{2} \right) \left( 1-\frac{\epsilon}{2} \right) =
\frac{\mu^{\prime *}_{L\tau}}{\sqrt{M_D}} .
\end{array}
\end{equation}

Finally, for completeness, we mention that the corresponding change in Eq.~(\ref{deltamass})
is with
\begin{equation}
\delta m_\nu = \epsilon \; \frac{m_2}{6}  \left(
\begin{array}{rcrcr}
0 && 1 && -1 \\ 1 && 2 && 0 \\ -1 && 0 && -2
\end{array} \right) +
\epsilon \; \frac{m_3}{2}  \left(
\begin{array}{rcrcr}
0 && -2 && 2 \\ -2 && 1 && 0 \\ 2 && 0 && -1
\end{array} \right) .
\end{equation}

It is then seen that the understanding of the mixing pattern lies in the flavor
symmetry defined in Eq.~(\ref{ansatz}) and that the symmetry breaking effects give
the corrections needed for a better description of flavor mixings.

Within the conventional parametrization of the unitary matrix given in Eqs.~(\ref{usedU})
and (\ref{Rmatrices}), and in the context of the discrete non-abelian symmetry groups
mentioned above, corrections on the TBM mixing are described in terms of new scalar fields,
the so-called flavon fields. These are neutral under the SM interactions but form nontrivial
representations of the symmetry flavor group. They produce the spontaneous breaking of the
discrete symmetry by acquiring proper vacuum expectation value alignments.\cite{TBMflavons}
For corrections with less emphasis on symmetry groups, see
Refs.~\refcite{TBMchanges1}--\refcite{TBMchanges13}.

In conclusion, we have shown that the low-scale minimal linear seesaw model of neutrino
mass generation, proposed recently to motivate a scale for sterile neutrinos that goes
down to the GeV or even keV range, explains the observed neutrino mixing pattern in terms
of the tri-bimaximal mixing and its deviations from the symmetry between $\mu$ and $\tau$
flavors, taking the currently unknown CP violating phases at the zero level
with a real neutrino mixing matrix. In order to accommodate nonzero values of these phases
in the present framework, we should consider from the beginning a complex mixing matrix,
which is beyond the scope of this paper.

The lowness of the seesaw scale is related to the soft breaking of the $Z_2$ symmetry
between LH and RH neutrinos in the even state, and the extra sterile neutrinos in the
odd state. A sterile neutrino of quasi-Dirac type is then expected at or below the GeV
scale. It should be noted here that there are several other and, in particular, more
economical models that are consistent with the neutrino flavor oscillation data and at
the same time contains sub-GeV mass heavy sterile neutrinos. However, in these models
either RH Majorana neutrinos are considered or its extensions with extra sterile neutrinos,
which have no participation in the generation of active neutrino masses, are discussed.
They are different from our proposal both from the model as well as the phenomenology point
of view.

The super-heavy RH neutrinos, because of their heaviness, are essentially
decoupled from the low-scale phenomenology and thus their contributions to the light
neutrino mass matrix and flavor mixing become strongly suppressed. Yet, they can generate
the baryon asymmetry of the universe via high-scale leptogenesis, a mechanism where the
three RH neutrinos decay into the lepton doublet and the Higgs doublet by means of the
Yukawa interactions. These processes, assumed out-of-thermal-equilibrium in the early
universe, are lepton-number-violating as the RH neutrinos are their own antiparticles
and CP-violating due to the asymmetry between such a decay and its CP-conjugate process
involving the antiparticles.
Since the extra sterile neutrinos take care of the light neutrino masses and
therefore no mass hierarchy within the super-heavy RH neutrinos is required, our option
is the unflavored leptogenesis with no connection between low and high energy CP
violations.\cite{NoCP1}$^{\mbox{--}}$\cite{NoCP3} The RH neutrino masses are so large that
all the relevant Yukawa interactions leading to the final lepton asymmetry are unable to
distinguish among lepton flavors.\cite{Leptogenesis1}$^{\mbox{--}}$\cite{Leptogenesis4}
In our scenario, the leptogenesis and the neutrino mass generation work at different
scales, in contrast to the usual construction where they have a same scale without
involving the extra sterile neutrinos.

The puzzle of DM was not addressed by this paper. However, the Dirac pair of extra
sterile neutrinos, odd under the $Z_2$ symmetry, appears as a natural DM candidate,
behaving as freeze-in type of DM.\cite{freezein1,freezein2} The essential weakness
of the active-dark neutrino mixing is related to the soft breaking of $Z_2$ symmetry
and the very tiny breaking of unitarity in the active mixing matrix. A study of this
scenario will be presented elsewhere.

\section*{Acknowledgement}

This work was partially supported by Vicerrector\'{\i}a de Investigaci\'on,
Desarrollo e Innovaci\'on, Universidad de Santiago de Chile (USACH).

%%%%%%%%%%%%%%%%%%%%%%%%%%%%%%%%%%%%%%%%%%%%%%%%%%%

\end{document}